\documentclass[preprint,showpacs,amsmath,amssymb,floatfix,pre]{revtex4}
\usepackage{graphicx}
\begin{document}
\title{Ferromagnetic phases due to  competing short- and long-range interactions in the spin-S Blume-Capel model} 

\author{Octavio D. Rodriguez Salmon \footnote{Corresponding author.\\ E-mail address: octaviors@gmail.com}}
\affiliation{Universidade Federal do Amazonas, Departamento de F\'{\i}sica, 3000, Japiim, 69077-000, Manaus-AM, Brazil}

\author{Minos A. Neto } 
\affiliation{Universidade Federal do Amazonas, Departamento de F\'{\i}sica, 3000, Japiim, 69077-000, Manaus-AM, Brazil}

\author{Francisco Din\'{o}la Neto } 
\affiliation{Centro Universit\'{a}rio do Norte - UNINORTE\\Rua Huascar de Figueiredo, 290, Centro\\
69020-220, Manaus, AM, Brazil.}

\author{Denis Am\'{e}rico Murillo Pariona} 
\affiliation{Universidad Nacional Mayor de San Marcos, Facultad de Ciencias F\'{i}sicas, Av. Venezuela s/n,
Apartado Postal 14-0149, Lima-1, Per\'{u}.}

\author{Justo Rojas Tapia } 
\affiliation{Universidad Nacional Mayor de San Marcos, Facultad de Ciencias F\'{i}sicas, Av. Venezuela s/n,
Apartado Postal 14-0149, Lima-1, Per\'{u}.}

\date{\today}

\begin{abstract}
\noindent
We broaden the study of  the statistical physics of the spin-$S$ Blume-Capel model with ferromagnetic mean-field interactions $J$ in competition with short-range antiferromagnetic interactions $K$ in a linear chain in the thermodynamic limit. This work describes the critical behavior of the model when the $S$ takes a  half integer  and an  integer value. In both cases the phase diagrams exhibit  new ferromagnetic phases (for certain values of $K$) enclosed by branches emerging from the first-order frontiers of the pure ferromagnetic model. For finite temperatures the complex topologies were obtained by numerical minimization of the free energy.  

\vskip \baselineskip
%\vspace{2cm}
%\medskip


\end{abstract}
\maketitle
\newpage

\section{Introduction}
The study of  systems with competing short- and long-range interactions attracts Condensed Matter physicists due to the complex phenomena that can emerge. Infinite-range interactions together with  short-range ones have  already been  produced in laboratories. For instance,  Landig et al. \cite{landig} have experimentally realized this kind of competition for a bosonic model in an optical lattice observing the appearance of four distinct phases, namely, a superfluid, a supersolid, a Mott insulator and a charge density wave \cite{landigthesis}. This motivates the theoretical study of models with interactions on different length scales. \\\\
In what  Ising-like models concern, it is an old problem \cite{morg} that still maintains the attention due to its complexity\cite{belim}. An early work, called the Nagel-Kardar model, considers the Ising model with mean-field ferromagnetic interactions in competition with short-range antiferromagnetic interactions \cite{nagle,kardar}. In a system with mean-field interactions  each spin interacts with the others with the same strength, so it interacts equally with  the closer neighbor and with the furthest one. The short-range couplings were considered as nearest-neigbor interactions in $d=1$, as well as in $d=2$ dimensions.  In $d=1$, a linear chain contain spins with first-neighbor antiferromagnetic couplings with mean-field ferromagnetic ones. This competition produces a tricritical point in the  frontier that separates the ferromagnetic phase ($\bf F$) and the paramagnetic phase ($\bf P$) of the phase diagram. For $d=2$, this becomes more complex on account of  the appearing of the antiferromagnetic phase ($\bf AF$)\cite{bonner,kaufman, mukamel,cohen}. This model has also been used to test ensemble inequivalence \cite{campa,duv}, raising some controversial answers \cite{salinas1, salinas2}. \\\\

In this work we  complement the studies of the  Nagel-Kardar version of the spin-$S$ Blume-Capel Model \cite{salmon1, salmon2} by considering the particular cases where $S=5/2$ and $S=2$, in order to summarize the resulting topologies of the phase diagrams when the model assumes  half-integer and  integer values of $S$, respectively. The paper is organized as follows, in section II we present the Hamiltonian representing the model, a brief derivation of the free energy density, and we describe the theoretical fundations of the numerical procedure to be applied for the construction of the phase diagrams; in section III we analyze the ground states  for  $S=5/2$ and $S=2$; and  the phase diagrams at finite temperatures are discussed in section IV.

\vskip \baselineskip
\noindent
\section{Theoretical background}

We treat in this paper a  version of the  spin-$S$ Blume-Capel Model on a linear  chain, which can be  represented by the following Hamiltonian:

\begin{equation}
\label{hamiltonian}
{\cal H} = -\frac{J}{2N} ( \sum_{i=1}^{N} S_{i}  )^{2} + K \sum_{i=1}^{N}S_{i}S_{i+1} + D \sum_{i=1}^{N}S_{i}^{2},
\end{equation}
 \vskip \baselineskip
\noindent
where $\{ S_{i} \}$ are classical spin variables, such that $S_{i} = -S,-S+1,..,S-1,S$, with $i=1,2...,N-1,N$, where $N$ is  the total  number of spins on the chain. The first sum stands for the energy of the  mean-field ferromagnetic interactions, because $J>0$. So,  each spin $S_{i}$ interacts with the same strength $J$ with all the $N-1$ others, and also with itself.   The second sum is the energy due to the antiferromagnetic nearest-neighbor interactions ($K>0$) between the spins, and the third sum is the anisotropy term, typical in the Blume-Capel model. Therefore, there is a competition between the $\bf F$ and the $\bf AF$ order that the two first sums tend to establish, according to the strength of $K$ in comparison to $J$, together with the value of the anisotropy constant $D$. 

\vskip \baselineskip
\noindent
In order to study the equilibrium  Statistical Physics of this model and its criticality we need the expression of the  free energy as a function of the order parameters. In this case the magnetization per spin is the  relevant order parameter for finite temperatures, since the antiferromagnetic phase appears only in the ground state (for the linear chain). However, an exact expression of  the free energy can be obtained only in a few cases.  Fortunately,  the partition function of the Hamiltonian in Eq.(\ref{hamiltonian}) can be "exactly" determined by applying the Hubbbard-Stratonovich transformation \cite{hubbard}, the Transfer Matrix technique \cite{baxter} and then the Steepest descent method \cite{skmodel}, leading to the following expression for the free energy per spin in the thermodynamic limit ($N \to \infty$):

\begin{equation}
\label{fdensity}
f =   \frac{1}{2}  J m^{2} - \frac{1}{\beta} \log(\lambda_{max}), 
\end{equation}
\vskip \baselineskip
\noindent
where $\beta = 1/k_{B}T$ and  $T$ is the temperature.  $\lambda_{max}$ is the maximum eigenvalue of the transfer matrix $\bf M$ (which is symmetric), whose elements are $ M_{\mu,\nu}$$=e^{ \beta  Q(i,i+1)}$, with $Q(i,i+1) =  \frac{1}{2}J m (S_{i}+S_{i+1}) - K S_{i}S_{i+1} -  \frac{1}{2} D (S_{i}^{2}+S_{i+1}^{2})$, being $m$  the magnetization per spin that minimizes the function  $f$,  for given values of $k_{B}T/J$, $D/J$ and $K/J$, at the equilibrium.  For simplicity, it is convenient to work with the reduced variables $t = k_{B}T/J$, $d=D/J$ and $k=K/J$. A detailed derivation of the free energy through the steepest descent method was exposed  by the authors in references \cite{salmon1} and \cite{salmon2}. \\

For finite temperatures, the free energy density is a fundamental tool to  explore the evolution of the phase diagram in the $d-t$ plane, as $k$ increases from zero. Note that $k$ is the antiferromagnetic nearest-neighbor coupling (in units of $J$) that competes with the ferromagnetic mean-field interactions, thus it tends to destroy the ferromagnetic phase. Also, it is important to emphasize that  this phase diagram consists of frontiers separating the different phases determined by the order parameter $m$, which minimizes $f=f(m)$, for given values of $t$, $d$, and $k$. So, we use numerical minimization, because  the obtention of the maximum eigenvalue $\lambda_{max}$ of $\bf M$ is hard  by hand for matrix elements more than 3x3 (see the appendix in reference \cite{salmon1}). In this case the Power Iteration Method is a useful method because it gets directly the largest eigenvalue \cite{power}. \\
Currently, the frontiers lines  are divided in two types, namely, first-  and second-order lines. The main difference is that the order parameter suffers a jump discontinuity at  the first-order frontiers, whereas it is continuous at the second-order ones. Furthermore, the function $f(|m|)$ presents coexistence of minima at first-order lines, nevertheless, $f(|m|)$ presents one global minima at  second-order points. These criteria are taken into account for our algorithm, which scans the frontier points for given values of $t$, $d$ and $k$. To plot the frontiers and points of  the phase diagrams we  use the following symbols:   

\begin{itemize}

\item Continuous (second order) critical frontier: black continuous line;

\item First-order  frontier (line of coexistence): gray continuous line;

\item Tricritical point (point dividing the first- and the second-order of a frontier line): represented  by a black circle;

\item Ordered critical point (point where a first-order frontier ends): represented by a black asterisk;

\end{itemize}

\section{Ground states}
The main tool to find the ground state phases, i.e., the phases of the system at zero temperature, is the energy of the Hamiltonian. The energy density of the Hamiltonian given Eq.(\ref{hamiltonian}), must be minimized for given values of the reduced parameters $k$ and $d$.  For $S=5/2$, we found twelve phases that can appear in the ground state as shown in Figure 1. Note that each spin can take six values, $S_{i}=-5/2,-3/2,-1/2,1/2, 3/2, 5/2$. Also, in Figure 1 there are six ferromagnetic phases (parallel alignment), labeled as $\bf F_{1}$, $\bf F_{2}$, $\bf F_{3}$, $\bf F_{4}$, $\bf F_{5}$ and $\bf F_{6}$, and six antiferromagnetic phases (antiparallel alignment),  labeled as $\bf AF_{1}$, $\bf AF_{2}$, $\bf AF_{3}$, $\bf AF_{4}$, $\bf AF_{5}$ and $\bf AF_{6}$. Of course, these phases are degenerated, so Figure 1 shows only one configuration state for each one. In Table I is exhibited their respective energy densities and magnetizations. From this information we  plotted the frontiers of the  phase diagram in the $k-d$ plane as observed in Figure 2(a). All frontier lines are of first order, due to the fact that the first derivative of Hamiltonian  energy is discontinuous at the points belonging these lines. In Figure 2(b) we illustrate it by ploting the energy density ($e_{gs}$) of the ground state as function of $k$, for the particular value $d=0.45$. Note that the first derivative of $e_{gs}$ is not continiuous at the points where its curve crosses the three frontier lines shown in Figure 2(a). \\\\
Another aspects to highlight about Figure 2 is the fact that the twelve phases coexist at the point $(k,d)=(0.25,0.25)$ represented by the diamond. From Table I we realize that the twelve one have the same energy (equal to zero) for $k=0.25$ and $d=0.25$. Also, phases $\bf F_{4}$, $\bf F_{5}$,  $\bf F_{6}$, $\bf AF_{4}$, $\bf AF_{5}$, and $\bf AF_{6}$ can only appear at this point, so out of it, these phases do not minimize $e_{gs}$, for any value of $k$ and $d$. On the other hand, phase $\bf F_{2}$ is present only at the frontier separating phases $\bf F_{1}$ and $\bf F_{3}$, whereas phase $\bf AF_{2}$ is only at the frontier dividing $\bf AF_{1}$ and $\bf AF_{3}$. For instance, the first  frontier (that on the left) can be obtained by equating the energy densities of phases $\bf F_{1}$, $\bf F_{3}$ and $\bf F_{2}$, giving the linear equation $d = 1/2-k$ (see the energy expressions of Table I). The same is done for the other frontiers (see this procedure in detail in reference \cite{salmon2}). Accordingly, in Figure 2(a) we may see that at the straight segment of equation $d = 1/2-k$ (for $0 < k < 0.25$) phases  $\bf F_{1}$, $\bf F_{3}$ and $\bf F_{2}$ coexist. The straight segment on the right is described by the equation $d = 1/2-k$ (for $k > 0.25$), and there, phases  $\bf AF_{1}$, $\bf AF_{3}$ and $\bf AF_{2}$ coexist, whereas at the frontier in the middle ($k=0.25$) phases $\bf F_{1}$ and $\bf AF_{1}$ coexist, for $0 < d < 0.25 $, and  $\bf F_{3}$ and $\bf AF_{3}$, for $d > 0.25$. \\

 For $S=2$, the procedure of exploring the ground state phases is the same as described above. We point out that when $S=2$, each spin can take five values, $S_{i}=-2,-1,0,1,2$. In Figure 3 we exhibit the nine phases that can appear at $T=0$, namely, five ferromagnetic phases $\bf F_{1}$, $\bf F_{2}$, $\bf F_{3}$, $\bf F_{4}$,  three antiferromagnetic ones $\bf AF_{1}$, $\bf AF_{2}$, and $\bf AF_{3}$, and the phase $\bf P$, not ordered. This not ordered phases is of null  magnetization, because all spins take the zero value.  In Table II is shown their respective energy expressions and magnetizations.  The phase diagram shown in Figure 4 is topologically similar to that for the $S=5/2$ case (see Figure 2(a)). All frontier lines are of first order, and the nine phases coexist at the point represented by the triangle, whose coordinates are $(k,d)=(0.25,0.25)$. This is the only point where  phases $\bf F_{3}$, $\bf F_{4}$ and $\bf AF_{3}$ appear for $T=0$. The linear frontier on the left separates phases $\bf F_{1}$ and $\bf P$, and only at this line the phase $\bf F_{2}$ is present, so at this straight segment (where $0< k < 0.25$), phases $\bf F_{1}$, $\bf F_{2}$ and $\bf P$ coexist. Similarly, this happens for the frontier on the right (for $k > 0.25$), for phases $\bf AF_{1}$, $\bf AF_{2}$ and $\bf P$, as illustrated in Figure 4. \\

The knowledge of the ground state is very important for the understanding of the behavior of the system at finite temperatures. Some imperceptible phases at $T=0$ (those who exist just at one point), will increase their extensions in the phase diagrams for $T>0$. We will show this in the next section.   

\vskip 2\baselineskip

\section{Phase Diagrams at finite temperatures}

The first effect of an infinitesimal value of temperature on this system is the disappearing of all  antiparallel spin configuration. Accordingly, the $\bf AF$ phases shown in the diagrams of Figure 2(a) and Figure 4 turn into the $\bf P$ phase for $T>0$. This happens because the interactions in the second sum of the Hamiltonian given in Eq.(\ref{hamiltonian}) are in one dimension (1D). So, in this case we simply have ferromagnetic phases for $0< k < 0.25$, then the only the phase $\bf P$ remains (for $k > 0.25$). However, the phase diagrams exhibit  an interesting evolution in the region where $0< k < 0.25$, while the region enclosing the ferromagnetic orderings is being reduced as $k$ increases. \\

We firstly analyze the evolution of the critical behavior of the system for the $S=5/2$ case, then the $S=2$ case presents a similar behavior. The description of the critical behavior of the system will be done through the information of the phase diagrams of the model in the $d-t$ plane, for different values of $k$. The main tool for obtaining the phase diagrams is the free energy density $f$ given in Eq.(\ref{fdensity}). An equilibrium state for given values of $d$, $k$ and $t$, is obtained by finding numerically the value of the order parameter  $m$ (the magnetization per spin), which minimizes the  function $f$. Thus, each phase point  is identified in the $d-t$ plane, according to its magnetization value. Nevertheless, a first-order point may have more than two coexisting magnetizations that equally minimize the free energy density. As commented  before, the magnetization is discontinuous in that case.\\

In Figure 5(a) is shown  the phase diagram of the Blume-Capel model in the $d-t$ plane, for $k=0$. There the most complex region is exhibited in the interval $0.492 < d < 0.504$. We see that the frontier dividing the ferromagnetic and   the paramagnetic regions is wholly of second order. For $d=0$, its critical temperature is $t_{c} \simeq 2.917$. We observe in the lower temperature region two first-order frontiers separating phases $\bf F_{1}$, $\bf F_{2}$ and $\bf F_{3}$. These lines finish at ordered critical points represented by  asterisks. In the vicinity of one of these ordered critical points, one can choose a pathway so as to  go continuously from one ferromagnetic phase to another. On the other hand, one may observe the behavior of the magnetization curve when crossing the three frontiers of the phase diagram. To this end we scanned the magnetization per spin $m$ as a function of the temperature for $d=0.4995$, in order to  ensure that  $m$ curve will cross these three frontiers, as shown Figure 5(b). There we may observe that the magnetization suffers a jump discontinuity when crossing the two first-order frontiers, whereas it falls continuously to zero when passing through the second-oder frontier dividing the ferromagnetic and paramagnetic phases. Before that, it reaches a local maximum, which is something odd. \\

Another effect of the short-range interaction $k$ is the gradual disappearing of second-order critical points. This has been already observed \cite{salmon1, salmon2}. So, the original second-order frontier is transformed into a first-order one as $k$ increases from a certain critical value. In Figure 6 is shown the evolution of this frontier for differente values of $k$. Note that for an interval in $k$, tricritical points appear so as to limit the extension of the second-order frontier, which is gradually reduced. Then, after some critical value of $k$ (in this case between $k=0.20$  and $k= 0.21 $), the extremes of the second-oder frontier meet themselves, then it  disappears, so the frontier becomes wholly of first order. \\

In Figure 7 we analyze the phase diagram for $k=0.20$. In Figure 7(a) the higher frontier dividing phases $\bf F$ and $\bf P$ is partially of second order, in agreement with what was shown in Figure 6. The vertical arrows at $d=0.10$, $d=0.25$ and $d=0.40$ are guides to the eye so as to see where the magnetization curve is plotted in Figure 7(b). These three magnetization curves confirm the nature of each section of the frontier line dividing  $\bf F$ phases  and the phase $\bf P$. We may observe that only the curve for $d=0.25$ is continuous when crossing the frontier, signaling a second-order transition. Furthermore, Figure 7(a) have an inset showing the zoomed small region in the interval $ 0.296 < d < 0.301$, which can't be well observed in the scale of the main figure. There is observed the appearing of the $\bf F_{4}$ (see Figure 1) phase in a small region enclosed by the frontier dividing $\bf F_{1}$ and $\bf F_{2}$ phases and a short emerging branch that also ends at an ordered critical point.  We recall that phase $\bf F_{4}$ appeared only at one point in the ground state, as shown in Figure 2(a). \\

The most interesting evolution of the phase diagram is shown for $k=0.24$ in Figure 8. We may see that phase $\bf F_{4}$ is now in a increased region because the emerging branch shown in the inset of Figure 7(a) has grown. Furthermore, it has emerged another branch from the frontier dividing phases $\bf F_{2}$ and $\bf F_{3}$ enclosing phase $\bf F_{6}$, which also existed only at one point in the phase diagram at $T=0$ (see Figures 1 and 2). In Figure 9(a) we zoomed the most complex region of Figure 8, showing a vertical arrow as a guide to the eye so as to mark where the magnetization is plotted in Figure 9(b), which is at $d=2591$. We choose this  value of $d$, because from it the curve $m$ versus $t$ crosses all the exhibited phases. So, in Figure 9(b) is shown the magnetization values of the ferromagnetic phases in the following  sequence: $\bf F_{1}$, $\bf F_{4}$,$\bf F_{2}$, $\bf F_{6}$ and $\bf F_{3}$, forming  an artimetic progression $2.5,2.0,1.5,1.0,0.5$, respectively.  Note that phases $\bf F_{4}$ and $\bf F_{6}$ have magnetizations values  $m=2$ and $m=1$ (at the frontier lines), respectively, the same as the ground state phases $\bf F_{4}$ and $\bf F_{6}$ (see Figure 1 and Table I). This strongly suggests that they are the same phases. Nevertheless, we can't know throught the free energy density if phase $\bf F_{5}$ could have emerged, because it has the same magnetization as $\bf F_{2}$ (see Table I), so only through a Monte Carlo Simulation we could see if that configuration appears at finite temperatures. \\

In Figure 10(a), the phase diagram is shown for $k=0.2486$, very close to the end of all ferromagnetic order (for $k > 0.25$), as shown by the ground-state diagram of Figure 1. The ferromagnetic has been considerably  reduced, and one may see the formation of closed ferromagnetic regions, due to the fact that the ordered critical points of the lower first-order frontiers are being "connected" to the higher frontier that limits the phase $\bf P$, as they were like frontier "nodes". To see it, we enclosed one of these frontier "nodes" by a circle, as a guide to the eye. That node has coordinates  $(d,t) = (0.0918713(5),0.248933(5))$, and we may see in Figure 10(b) which phases coexist in it. We observe that the magnetizations of phases $\bf F_{2}$, $\bf F_{6}$ and $\bf P$ minimize equally the free energy. The same behavior was observed in reference \cite{salmon2}. Therefore, this is essentially the critical behavior for the $S=5/2$ case, which may represent well when $S$ assumes half-integer values. \\

For the sake of completeness, we treat briefly the $S=2$ case, which can represent what happens with the model when $S$ assumes integer values. So, in Figures 11(a) and 11(b) we show how the topology of the phase diagram of the spin-$2$ Blume Capel model is transformed for $k=0.24$. Initially, the pure mean-field model (for $k=0$) shows a frontier separating the ferromagnetic zone and the paramagnetic phase $\bf P$. This frontier line has a second-order section and a first-order one divided by a tricritical point. For greater values of $k$ this frontier will be wholly of first order (as observed for $S=5/2$). A lower first-order line divides two ferromagnetic phases $\bf F_{1}$ and $\bf F_{2}$ (the same included in the ground state described in Figure \ref{figura3}). In Figure 11(b) (for $k=0.24$) we note that all second order criticality has disappeared, and two  ferromagnetic phases $\bf F_{3}$ and $\bf F_{5}$ have emerged, most probably from the ground state (see Figures 3 and 4 and Table II). This happened on account of the emergence of  two new branches (two new first-order frontier lines) each sprouted from the two first-order frontiers of the phase diagram obtained for $k=0$. In order to confirm the magnetization values of these phases we plotted the curve $m$ versus $t$, conveniently for $d=0.2575$. It is just the same analysis done in Figure 9 (for the $S=5/2$ case).  However, as happened with phase $\bf F_{5}$ when $S=5/2$,  we can't know if the ground-state phase $\bf F_{4}$ could have emerged because it has the same magnetization (though different configuration) as $\bf F_{2}$ (see Table II and the configurations shown in Figure 3). As mentioned before, the free energy density given in Eq.(\ref{fdensity}) can't distinguish spin configurations having the same $m$. Nevertheless, that information might be provided by a Metropolis replica exchange simulation \cite{parallel}. \\

Finally, in Figure 13 the phase diagram (for $S=2$) was obtained when $k=0.2485$, very close to $k=0.25$, before the end of the ferromagnetic order (see Figure 4). Although the topology is something different of that of Figure 10(a), some ferromagnetic regions are now enclosed too. These closed frontier lines contain some "nodes" of multiple coexistence, as demonstrated by the free energy density in Figure 10(b). Accordingly, this phenomenom happens for half-integer values of $S$ as well as for integer ones.

{\large\bf Acknowledgments}

Financial support from CNPq and FAPEAM (Brazilian agencies) is acknowledged. 

\vskip 2\baselineskip

%%%%%%%%%%%%%%%%%%%

\begin{table}[h!]
\centering
 \begin{tabular}{||c | c | c ||} 
 \hline
 Phase & Energy density & $|m|$ \\ [0.5ex] 
 \hline\hline
 $\bf F_{1}$ & $-\frac{25}{8}+\frac{25}{4}k+\frac{25}{4}d$ & 5/2 \\ 
 \hline
 $\bf F_{2}$ & -$\frac{9}{8}+\frac{9}{4}k+\frac{9}{4}d$ & 3/2 \\ 
 \hline
 $\bf F_{3}$ &  $-\frac{1}{8}+\frac{1}{4}k+\frac{1}{4}d$ & 1/2 \\ 
 \hline
 $\bf F_{4}$ & $-2+\frac{15}{4}k+\frac{17}{4}d$ & 2 \\ 
 \hline
  $\bf F_{5}$ & $-\frac{9}{8}+\frac{5}{4}k+\frac{13}{4}d$ & 3/2 \\
 \hline 
 $\bf F_{6}$ & $-\frac{1}{2}+\frac{3}{4}k+\frac{5}{4}d$ & 1 \\ 
 \hline
 $\bf AF_{1}$ & $-\frac{25}{4}k+\frac{25}{4}d$ & 0 \\ 
 \hline
 $\bf AF_{2}$ & $-\frac{9}{4}k+\frac{9}{4}d$ & 0 \\ 
 \hline
 $\bf AF_{3}$ & $-\frac{1}{4}k+\frac{1}{4}d$ & 0 \\ 
 \hline
 $\bf AF_{4}$ & $-\frac{1}{8}-\frac{15}{4}k+\frac{17}{4}d$ & 1/2 \\ 
 \hline
  $\bf AF_{5}$ & $-\frac{1}{2}-\frac{5}{4}k+\frac{13}{4}d$ & 1 \\
 \hline
  $\bf AF_{6}$  & $-\frac{1}{8}-\frac{3}{4}k+\frac{5}{4}d$ & 1/2 \\ [1ex] 
 \hline
 \end{tabular}
 \caption{For $S=5/2$ : phases appearing in the ground state with their respective energy densities  in $J$ units (the energy density corresponds to the Hamiltonian given in  Eq.(\ref{hamiltonian})), with the corresponding absolute value of the magnetization $m$. The values of these energy expressions are global minima of the  Hamiltonian energy density just for certain values of $k$ and $d$.}
\end{table}

%%%%%%%%%%%%%%%%%%%%%%%%%%%

%%%%%%%%%%%%%%%%%%%

\begin{table}[h!]
\centering
 \begin{tabular}{||c | c | c ||} 
 \hline
 Phase & Energy density & $|m|$ \\ [0.5ex] 
 \hline\hline
 $\bf F_{1}$ & $-2+4k+4d$ & 2 \\ 
 \hline
 $\bf F_{2}$ & -$\frac{1}{2}+k+d$ & 1 \\ 
 \hline
 $\bf F_{3}$ &  $-\frac{9}{8}+2k+\frac{5}{2}d$ & 3/2 \\ 
 \hline
 $\bf F_{4}$ & -$\frac{1}{2}+2d$ & 1 \\ 
 \hline
 $\bf F_{5}$ &  $-\frac{1}{8}+\frac{1}{2}d$ & 1/2 \\ 
 \hline
$\bf AF_{1}$  & $-4k+4d$ & 0 \\
 \hline 
 $\bf AF_{2}$  & $-k+d$ & 0 \\
 \hline 
   $\bf AF_{3}$  & $-\frac{1}{8}-2k+\frac{5}{2}d$ & 1 \\ 
  \hline
  $\bf P$  & 0 & 0 \\ [1ex] 
 \hline
 \end{tabular}
 \caption{Energy density of the Hamiltonian in $J$ units (see Eq.(\ref{hamiltonian})), with the corresponding absolute value of the magnetization $m$, for $S=2$.}
\end{table}

%%%%%%%%%%%%%%%%%%%%%%%%%%%

%%%%%%%%%%%%%%%%%%%%%%%%%%%%
\vskip \baselineskip
\vspace{3.0 cm}
\begin{figure}[htbp]
\centering
\includegraphics[height=5cm]{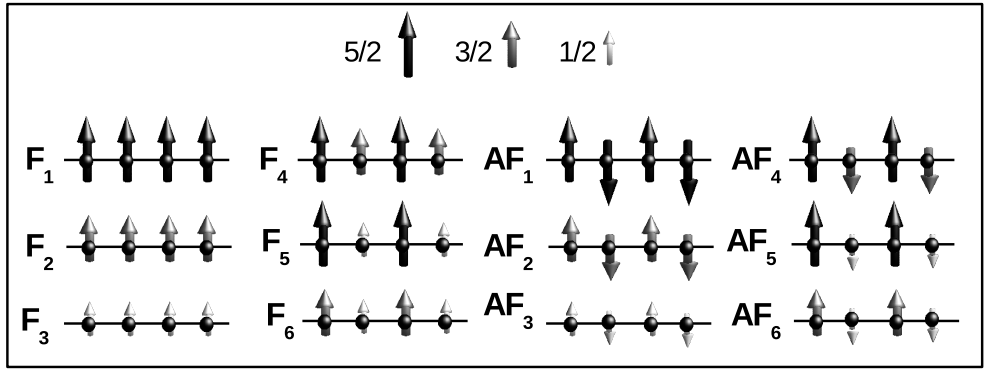}
\caption{Scheme of the representative spin configurations of the ground-state phases, for $S=5/2$. } 
\label{figura1}
\end{figure}
%%%%%%%%%%%%%%%%%%%%%%%%%%%%

%%%%%%%%%%%%%%%%%%%%%%%%%%%%
\vskip \baselineskip
\vspace{5.0 cm}
\begin{figure}[htp]
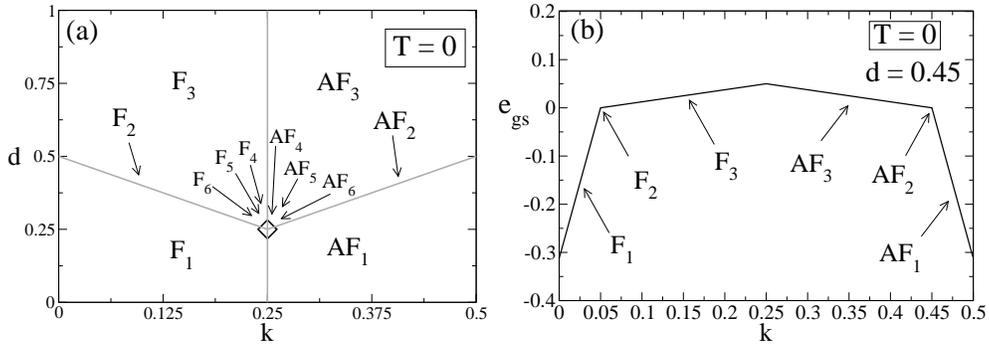

\centering
\includegraphics[height=4.5cm]{Figure2a.eps}
\includegraphics[height=4.5cm]{Figure2b.eps}
\caption{In (a) is shown  the phase diagram of the model at zero temperature, for $S=5/2$. Twelve phases coexist at the point represented by the diamond.  The frontier lines are of first order as demonstrated in (b). In (b) we exhibit the energy versus $k$, for $d=0.45$. It can be  seen that this curve suffers a discontinuity of its first derivative when crossing the frontier lines (shown in (a)). Thus,  at those points the transition is of first order. } 
\label{figura2}
\end{figure}
%%%%%%%%%%%%%%%%%%%%%%%%%%%%

%%%%%%%%%%%%%%%%%%%%%%%%%%%%
\vskip \baselineskip
\vspace{3.0 cm}
\begin{figure}[htbp]
\centering
\includegraphics[height=5cm]{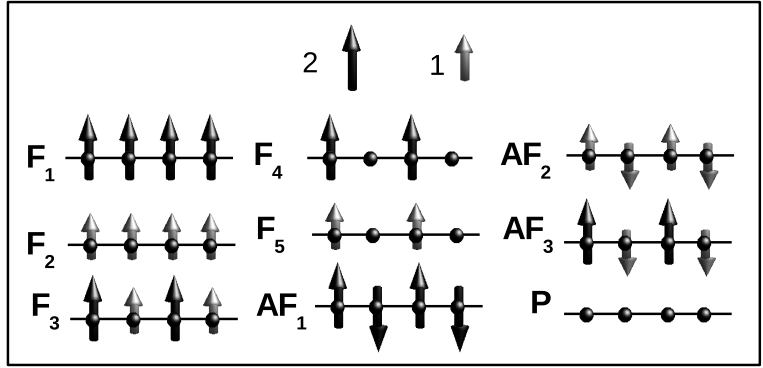}
\caption{Scheme of the representative spin configurations of the ground-state phases, for $S=2$. The sites without spins have $S_{i}=0$, as for the $\bf F_{4}$ and $\bf F_{5}$ phases. } 
\label{figura3}
\end{figure}
%%%%%%%%%%%%%%%%%%%%%%%%%%%%

%%%%%%%%%%%%%%%%%%%%%%%%%%%%
\vskip \baselineskip
\vspace{3.0 cm}
\begin{figure}[htbp]
\centering
\includegraphics[height=5cm]{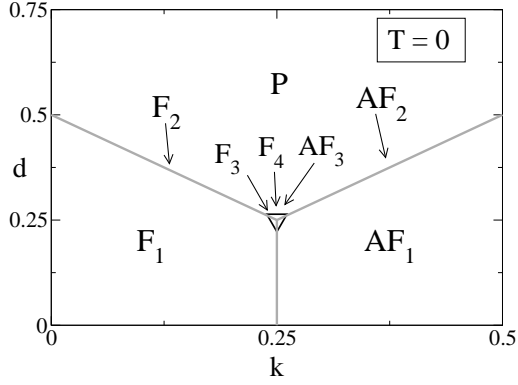}
\caption{The phase diagram of the model at zero temperature, for $S=2$. Eight phases coexist at the point represented by the triangle.} 
\label{figura4}
\end{figure}
%%%%%%%%%%%%%%%%%%%%%%%%%%%%

%%%%%%%%%%%%%%%%%%%%%%%%%%%%
\vskip \baselineskip
\vspace{3.0 cm}
\begin{figure}[htbp]
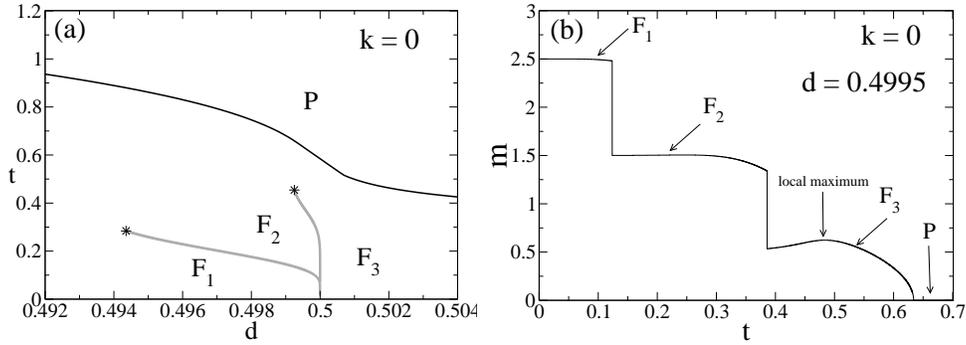

\centering
\includegraphics[height=4.5cm]{Figure5a.eps}
\includegraphics[height=4.5cm]{Figure5b.eps}
\caption{Phase diagram of the spin-$5/2$ ferromagnetic Blume-Capel Model in the $d-t$ plane, where $d$ and $t$ are reduced variables for the anisotropy and the temperature, respectively. We recall that the black continuous line is a second-order frontier, whereas the gray frontiers below are of first order. The asterisks are ordered critical points (where the first-order curves finish).  In (a) the most complex portion of the phase diagram; in (b) it is shown the magnetization curve crossing the most complex region of the frontiers (for $d$ close to 0.5).} 
\label{figura5}
\end{figure}
%%%%%%%%%%%%%%%%%%%%%%%%%%%%

%%%%%%%%%%%%%%%%%%%%%%%%%%%%
\vskip \baselineskip
\vspace{3.0 cm}
\begin{figure}[htbp]
\centering
\includegraphics[height=6cm]{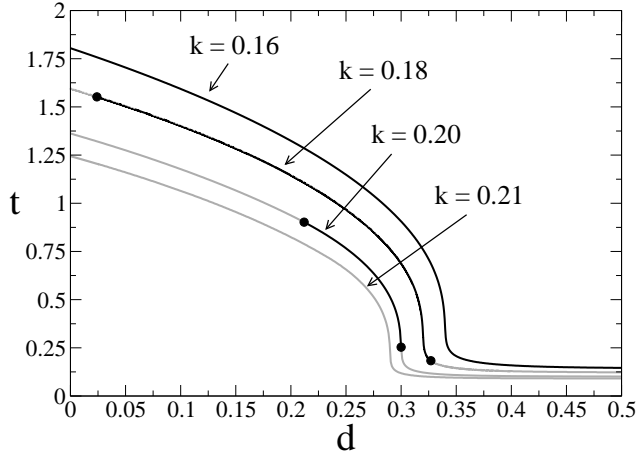}
\caption{This figure exhibits only the frontier line dividing the ferromagnetic and paramagnetic phases, for different values of $k$ (for $S=5/2$). The black circles are tricritical points. } 
\label{figura6}
\end{figure}
%%%%%%%%%%%%%%%%%%%%%%%%%%%%

%%%%%%%%%%%%%%%%%%%%%%%%%%%%
\vskip \baselineskip
\vspace{3.0 cm}
\begin{figure}[htbp]
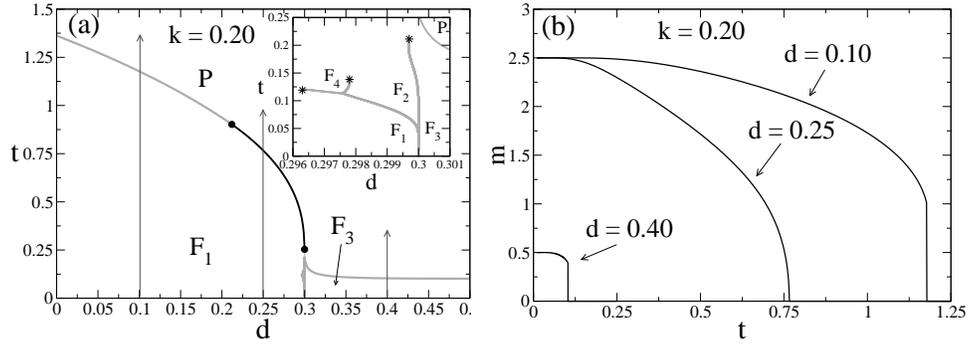

\centering
\includegraphics[height=4.5cm]{Figure7a.eps}
\includegraphics[height=4.5cm]{Figure7b.eps}
\caption{In (a) is shown the phase diagram for $k=0.20$. The inset shows the zoomed  lower region around $d=0.30$. The arrows are guides to the eye in order to show where the magnetization is scanned in (b). In (b) is presented the magnetization curves for three values of $d$, as indicated by the arrows in (a). } 
\label{figura7}
\end{figure}
%%%%%%%%%%%%%%%%%%%%%%%%%%%%

%%%%%%%%%%%%%%%%%%%%%%%%%%%%
\vskip \baselineskip
\vspace{3.0 cm}
\begin{figure}[htbp]
\centering
\includegraphics[height=6cm]{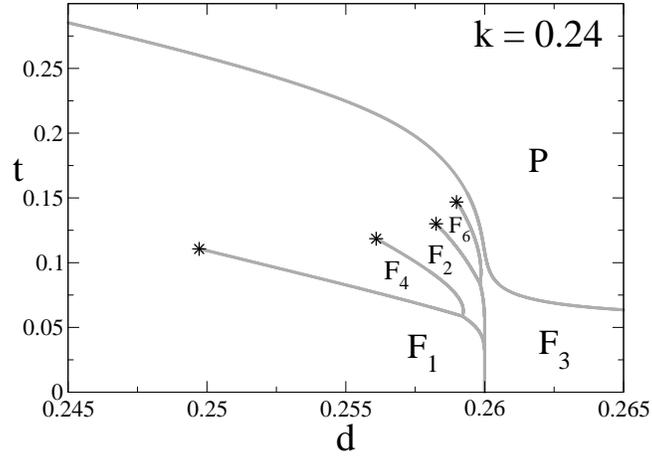}
\caption{The phase diagram for $k=0.24$, which is a representative value, by which we may observe the growing first-order branches that have emerged from the lower first-order lines of the pure ferromagnetic model (see Figure 1(a)).} 
\label{figura8}
\end{figure}
%%%%%%%%%%%%%%%%%%%%%%%%%%%%

%%%%%%%%%%%%%%%%%%%%%%%%%%%%
\vskip \baselineskip
\vspace{3.0 cm}
\begin{figure}[htbp]
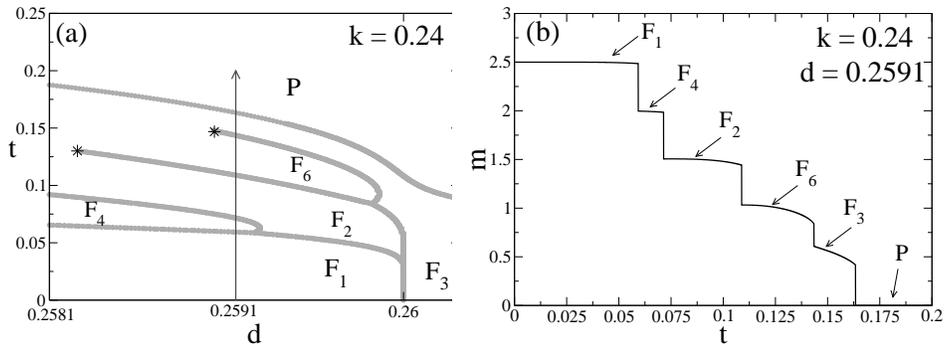

\centering
\includegraphics[height=4.5cm]{Figure9a.eps}
\includegraphics[height=4.5cm]{Figure9b.eps}
\caption{In (a) is exhibited a zoomed a region of Figure \ref{figura8} so as to show  where the magnetization is plotted in (b) as indicated by the arrow (a guide to the eye). In (b) is shown the magnetization curve in the direction indicated by the arrow in (a).  } 
\label{figura9}
\end{figure}
%%%%%%%%%%%%%%%%%%%%%%%%%%%%

%%%%%%%%%%%%%%%%%%%%%%%%%%%%
\vskip \baselineskip
\vspace{3.0 cm}
\begin{figure}[htbp]
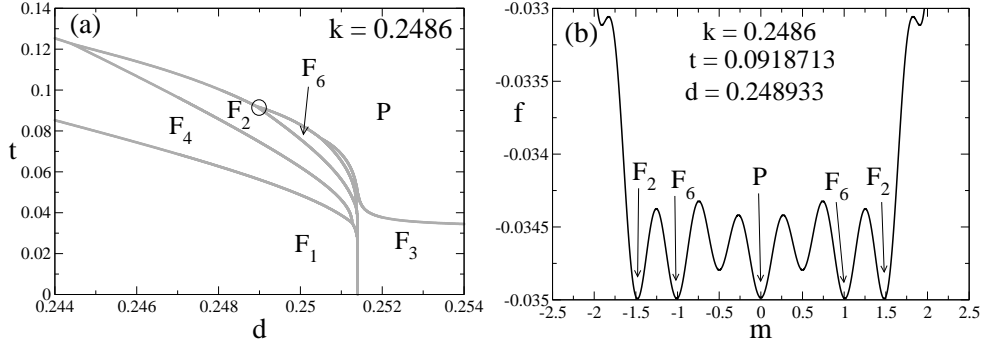

\centering
\includegraphics[height=4.5cm]{Figure10a.eps}
\includegraphics[height=4.5cm]{Figure10b.eps}
\caption{In (a) is shown the phase diagram of the model for a $k$ value close to $k=0.25$, just before the destruction of all ferromagnetic order. The point surrounded by the circle, which is the intersection of three first-order frontiers,  is studied by the free energy density in (b). In (b) we see the free energy density (in $J$ units) as a function of the magnetization per spin at the point surrounded by the circle in (a).} 
\label{figura10}
\end{figure}
%%%%%%%%%%%%%%%%%%%%%%%%%%%%

%%%%%%%%%%%%%%%%%%%%%%%%%%%%
\vskip \baselineskip
\vspace{3.0 cm}
\begin{figure}[htbp]
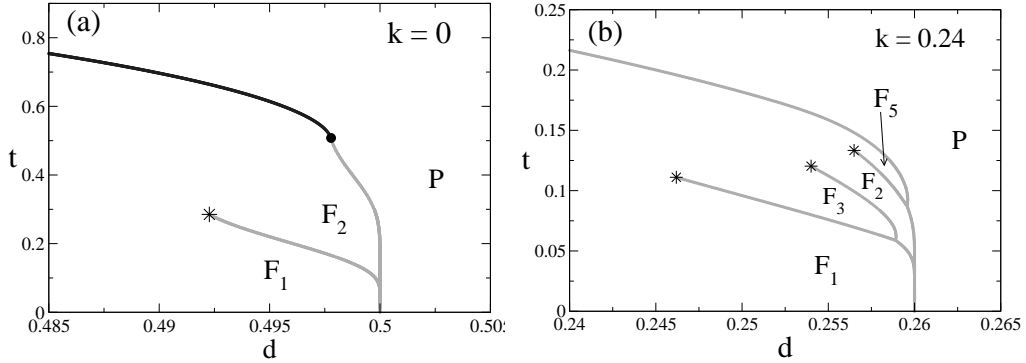

\centering
\includegraphics[height=4.75cm]{Figure11a.eps}
\includegraphics[height=4.75cm]{Figure11b.eps}
\caption{ In (a) is shown a portion of the phase diagram of the spin-$2$ Blume-Capel model with only mean-field ferromagnetic interactions ($k=0$). The most complex zone is exhibited. For $d=0$, $t_{c}=2.0$. In (b) is presented the evolved  phase diagram when $k=0.24$ (see the Hamiltonian written in Eq.(\ref{hamiltonian}). } 
\label{figura11}
\end{figure}
%%%%%%%%%%%%%%%%%%%%%%%%%%%%

%%%%%%%%%%%%%%%%%%%%%%%%%%%%
\vskip \baselineskip
\vspace{3.0 cm}
\begin{figure}[htbp]
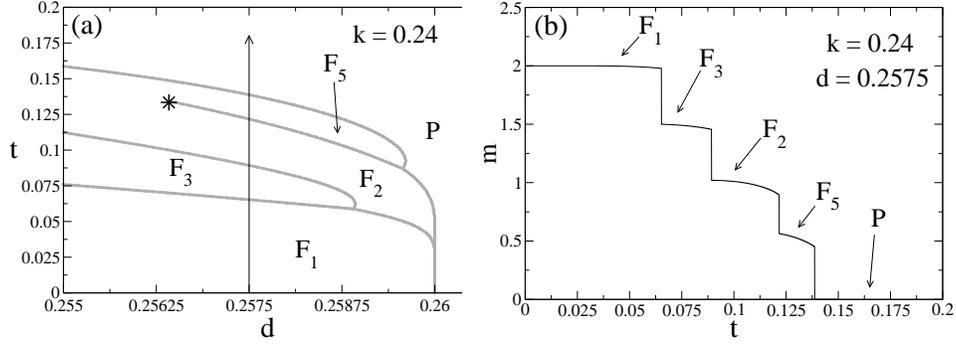

\centering
\includegraphics[height=4.5cm]{Figure12a.eps}
\includegraphics[height=4.5cm]{Figure12b.eps}
\caption{In (a) it has been   zoomed a region of Figure \ref{figura11}(b), where the arrow (a guide to the eye) follows  the pathway where the magnetization is plotted in (b). In (b) is shown the magnetization curve following  the direction indicated by the arrow in (a).  } 
\label{figura12}
\end{figure}
%%%%%%%%%%%%%%%%%%%%%%%%%%%

%%%%%%%%%%%%%%%%%%%%%%%%%%%%
\vskip \baselineskip
\vspace{3.0 cm}
\begin{figure}[htbp]
\centering
\includegraphics[height=6cm]{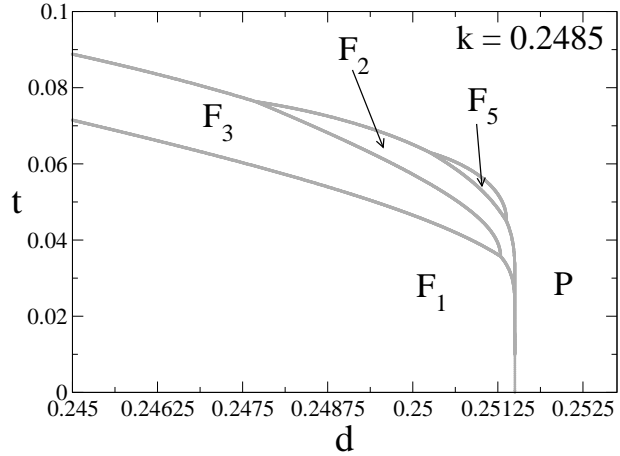}
\caption{A portion of the phase diagram of the model for $S=2$, for $k=0.2485$. We may observe that the open ferromagnetic regions (see Figure \ref{figura11}(b)) are being closed for the first-order frontiers.} 
\label{figura13}
\end{figure}
%%%%%%%%%%%%%%%%%%%%%%%%%%%%

\end{document}